# Why more contact may increase cultural polarization

Andreas Flache, University of Groningen, The Netherlands. email: a.flache@rug.nl

Michael W. Macy, Cornell University, email: mwm14@cornell.edu

## Abstract

Following Axelrod's model of cultural dissemination, formal computational studies of cultural influence have suggested that more possibilities of contact between geographically distant regions – for example due to improved communication technologies - may increase cultural homogeneity and decrease polarization of the overall society. However, more recent extensions point to the possibility that more contact may also sustain diversity by allowing local minorities to find outside support against pressures to conform to local majorities. In the present paper, we argue that these studies may not have gone far enough. We show that two relatively small but plausible and consequential extensions of Axelrod's social influence mechanism turn the effect of range of communication upside-down. The first extension is to assume a continuous opinion space and thus align Axelrod's model with previous studies of social influence. The second is to add the empirically well supported micro mechanisms of rejection and heterophobia to the model. Computational analyses of the resulting model demonstrate that now a larger range of contact can increase rather than decreases the extent of polarization in the population, contrary to Axelrod's original conclusion. Further experiments identify the window of conditions under which the effect obtains.

**Why more contact may increase cultural polarization[1]**

**Andreas Flache, University of Groningen, The Netherlands**

**Michael W. Macy, Cornell University, email: mwm14@cornell.edu**

1. Introduction

Recent decades have seen a dramatic increase in the possibilities for contact between people from geographically distant regions of the world, both due to more migration and better communication technologies (email, internet). Following Axelrod (1997), a range of formal models of cultural dynamics have been proposed that allow to address by what mechanisms and under what conditions this development may affect cultural diversity (e.g. Mark 1998, 2003; Klemm et al 2003a,b; Kitts et al 1999; Latané and Nowak, 1997). Most of these models combine two elementary social mechanisms, homophily and social influence. *Homophily* is attraction towards similar others, or the law that "birds of a feather flock together" (McPherson et al, 2001). *Social influence* is the tendency to adopt opinions of influential others (Festinger et al, 1950). In models of cultural dynamics, these two mechanisms usually are linked to each other by the assumptions that others are more influential if they are more similar to the focal actor and that social influence in turn increases the similarity of interactants.

A number of studies supported the view that more contact increases cultural homogeneity. Axelrod (1997) operationalized contact in terms of the overlap of neighborhoods in a cellular grid. If two agents are neighbors and they agree on at least one "feature", or issue, then there is positive probability that they will also come to agree on other issues (cf. Mark, 1998). In this model, cultural diversity can be a stable pattern despite ongoing social influence to adapt to neighbors' cultures, because influence stops when actors disagree in all dimensions of the opinion space. However, diversity is much reduced and uniform cultures tend to emerge when the geographical range within which neighbors influence one another increases in size. The reason is that a larger range of interaction exposes actors in different local regions to increasingly similar influences. As a consequence, it becomes unlikely that local regions are so dissimilar from their neighbors that they are entirely cut off from outside influences. Follow-up studies (e.g. Parisi et al 2003) somewhat modify the exact conditions, but retain the main result.

[1] This work is embedded in collaboration with Michael W. Macy (Cornell), James Kitts (Washington State), Karoly Takacs and Michael Maes (both University of Groningen), whose insightful comments and criticism is gratefully acknowledged. The research has been supported by the Netherlands Organization for Scientific Research (NWO) under the Innovational Research Incentives Scheme (VIDI).

Other extensions of Axelrod's model (e.g. Greig, 2002; Shibani et al 2001) point to a different possibility. They demonstrated that a more global range of interaction may allow local minorities to find in global communications support against pressures to conform to local majorities. When this happens, local minorities may no longer assimilate to the cultural profile of the local majority. In this view, more contact stabilizes diversity.

We agree with the view that more contact does not inevitably entail less diversity. However, we believe previous critics may still have underestimated the effect. A larger geographical range of contacts may not just help to retain diversity, but it may actually radically increase the differences between cultures. We will show in this paper that two relatively small but plausible and consequential modifications of Axelrods' original model suffice to obtain this implication.

The first modification addresses the conceptualization of culture and polarization. Axelrod (1997) used a conceptualization of polarization that obscures the fundamental difference between stable diversity and antagonism. A "polarized" outcome is seen as equivalent to a multicultural state in which neighboring cultures are different on all issues (or features). Cultural forms or opinions are modeled as nominal scales. Hence, agents can differ in opinion but the extent of their difference can not be expressed in the model. However, many cultural issues are better described by ordered or even continuous scales. Most observers would agree that the difference between Jazz and Blues is smaller than the difference between any two of these and Disco. Accordingly, it is well possible that a "polarized" state in Axelrod's terms is one in which a number of cultures are only slightly different from each other, but they are different on all aspects. This is clearly inconsistent with empirical studies of polarization (e.g. DiMaggio, Evans, and Bryson 1996), who see polarization as the tendency to hold an extreme and uncompromising view. More importantly, maximal disagreement, the only condition under which two neighbors stop to influence each other according to these models, is a much more exceptional and extreme state when issues are continuous. It requires that two neighbors adopt opposite ends of the opinion scales on all dimensions of the opinion space. Accordingly, we believe that previous modeling work may have overestimated how much cultural diversity can be generated by homophily and social influence alone. To test this, we integrate Axelrod's model with an older line of social influence models that neglected homophily, but did use continuous opinion scales (e.g. Abelson 1964, Friedkin and Johnson, 1999).

Moving towards a more fine grained representation of opinion scales seems to make the task of explaining stable diversity more difficult. What then may generate persistent diversity or, polarization? To answer this question, we draw upon our previous work (Macy et al, 2003) and propose a second modification of earlier models. In most previous modeling studies, social influence can only be positive in the sense that in the course of interaction opinions become more similar. The only alternative in Axelrod's model is no interaction at all. However, recent modeling work (e.g. Deffuant et al, 2005; Flache and Torenvlied, 2004) has shown that outcomes of social influence processes may radically change when at least a minority of agents insulate themselves from outside influence, but exert at the same time influence on more moderate others. Under this assumption, a few extremists suffice to split the agent society into a small number of opposed factions (Deffuant et al, 2005; Hegselmann & Krause, 200?).

Empirical evidence suggests that social actors may not only isolate themselves from social influence, but they may be negatively influenced by others' opinions (see e.g. Hass, 1981; Schwartz and Ames, 1977; Mazen and Leventhal, 1972). Modeling work in the wake of Axelrod neglected this mechanism of *rejection* or distancing as the mirror image of positive social influence[2].

Formal models of cultural dynamics also neglected the negative mirror image of homophily, *heterophobia*. *Heterophobia* describes the negative evaluation of others who are dissimilar. There is some empirical support for the existence of heterophobia in field experiments and in school research, where it was found that contacts can entail segregation of friendship networks (Baerveldt and Snijders, 1994; Moody, 2001).

Following our earlier work (Macy et al 2003), we propose a model of cultural dynamics that integrates the empirically documented mechanisms of rejection and heterophobia. We have shown that this extension can radically change cultural dynamics so that polarization becomes the most likely and robust outcome of cultural dynamics under a large of conditions. However, this work relied on dichotomous opinion scales and did not change contact structures. In the present paper, we move further and explicitly model continuous opinions and effects of contact structures.

In the remainder of this paper our model will be explained in Section 2. Section 3 summarizes the analyses and results. A discussion and conclusions follow in Section 4.

2. Model specification

Our model is an extension of Hopfield's attractor network (Hopfield, 1982; Nowak & Vallacher, 1997). Each node has one or more binary or continuous *states* and is linked to other nodes through endogenous *weights*. Weights change over time through a Hebbian learning rule (Hebb, 1949): the weight $w_{ij}$ is a function of the similarity of states for nodes $i$ and $j$ over time. More technically, the opinion of agent $i$ on issue $k$, ($1 \leq k \leq K$), $s_{ik}$, ranges between -1 and +1, ($-1 \leq s_{ik} \leq$ 1). If agents *i* and *j* are neighbors, there is a directed positive or negative weight $w_{ij}$, ($-1 \leq w_{ij} \leq 1$), otherwise the weight is zero. Time is modeled in discreet influence round. Social influence is modeled such that whenever *i* adopts his opinions, the change in the position of *i* with regard to issue *s* is a weighted sum of the distances to all neighbors, $s_{jk}$- $s_{ik}$, where the weights are given by the positive and negative ties. When the tie is negative, the focal actor is influenced to move away from the source of influence, when the tie is positive, he moves towards the source of influence on the corresponding opinion dimension. Moreover, influence is modified by a degree of "moderation" *m*, that expresses the extent to which influence decreases when opinions differences get smaller.

---
[2] Mark (2003) is as an exception, but he did not integrate rejection with heterophobia, as we propose in the present paper. Moreover, Mark did not apply his distancing model to address effects of increased contact on diversity.

We see moderation as a society level property, capturing the extent of "tolerance" with regard to small cultural differences in a society. We take a moderation of *m*=1 as the smallest meaningful level. This corresponds to a linear influence function, as used by almost all previous models of opinion formation in continuous opinion spaces. Higher levels of moderation correspond to more "tolerance". Technically:

$$s_{i,t+1} = s_{i,t} + \frac{1}{N} \sum_{j \neq i} w_{ij} \left( \frac{s_{j,t} - s_{i,t}}{2} \right)^m \quad (1)$$

where *N* is the size of the local neighborhood and *j* and *i* are members of that neighborhood. Purely positive influence is modeled with the assumption of $0 \leq w \leq 1$, while for positive and negative influence $-1 \leq w \leq 1$. To be precise, the power in the influence function retains the sign of the distance of opinions, and we impose a correction to avoid that opinions go out of bounds. For this, equation (1) is implemented in two steps, by equations (1a) and (1b) as follows:

$$\Delta s_{i,t} = \frac{1}{N} \sum_{j \neq i} w_{ij} \, Sign(s_{j,t} - s_{i,t}) \left( \left| \frac{s_{j,t} - s_{i,t}}{2} \right| \right)^m, \text{ where } Sign(x) = \begin{cases} +1, if \ x > 0 \\ -1, if \ x < 0 \\ 0, otherwise \end{cases} \quad (1a)$$

$$s_{i,t+1} = \begin{cases} s_{i,t} + \Delta s_{i,t}(1 - s_{i,t}), \text{if } s_{i,t} > 0 \\ s_{i,t} + \Delta s_{i,t}(1 + s_{i,t}), \text{if } s_{i,t} \leq 0 \end{cases} \quad (1b)$$

For homophily and heterophobia, we assume that the weight $w_{ij}$ increases or decreases over time depending on agreement between *i* and *j* in the *K* opinions. If agreement exceeds a threshold level then influence is positive, otherwise it is negative. For simplicity, this threshold is the midpoint of the interval of opinion differences (zero). Equation (2) formalizes the weight dynamics.

$$w_{ij,t+1} = w_{ijt}(1-\lambda) + \lambda(1 - \frac{\sum_{k=1}^{K}|s_{ikt} - s_{jkt}|}{K}) \quad (2)$$

For -1 ≤ s ≤ +1 this function yields results in the range -1 ≤ w ≤ +1. The parameter λ specifies a structural learning rate, ranging between 0 and 1. The higher λ in this range, the faster weights move towards the present level of (dis)agreement. To distinguish pure homophily from combined positive and negative social selection, the second part of the sum in (2) is modified as follows:

For homophily only: $w_{ij,t+1} = w_{ijt}(1-\lambda) + \lambda(1 - \frac{\sum_{k=1}^{K}|s_{ikt} - s_{jkt}|}{2K}) \quad (2a)$

Social influence and social selection are channeled by an exogenously imposed contact structure. Like Axelrod (1997), we use a regular grid structure, but we simplify it towards a one-dimensional circular grid. Figure 1 below shows how contact range is manipulated. Contact range (*r*) is the population fraction of neighbors to the "left" and "right" of a focal agent who are his neighbors, where neighborhoods are symmetrical.

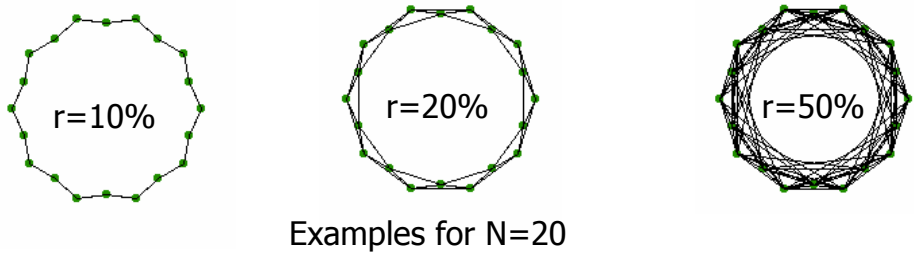

Examples for N=20

Figure 1: variation of contact range in regular grid structure

To avoid artifacts, we use asynchronous updating. In every period one agent is selected at random to update either a randomly selected state or a randomly selected weight with equal probability. The duration of a simulation run is expressed in number of iterations, where one iteration corresponds to $N*(K +$ average number of contacts) periods, to assure that on average every state and every weight is updated once within one iteration.

3. Results

We report four simulation experiments. The first two experiments test the effects of our model modifications using a baseline scenario comparable to previous studies. The third and fourth experiment address effects of contact range.

Experiment 1

Using a nominal opinion space, Axelrod (1997) found that diversity – measured in terms of the number of culturally distinct stable groups that co-exist in equilibrium – was fostered by a small number of features (dimensions) of the opinion space and a strongly local neighborhood, i.e. small overlap between the neighborhoods of neighboring positions on the grid. Accordingly, we made overlap between neighborhoods minimal in the sense that all agents had exactly one neighbor on both sides. Population size $N$ was set to N=100, yielding a contact range $r$=2%. The number of dimensions of the opinion space was likewise minimal with $k$=1. Further assumptions are a linear influence function ($m$=1), fast adaptation of weights ($\lambda$=1) and uniformly distributed initial opinions between -1..+1. To model homophily without heterophobia, initial weights are updated based on equation (2a), with a weight range between 0 and 1. Initial weights were set proportional to the level of initial agreement.

Figure 1 charts the change of two measures of diversity for the first 6000 iterations of a typical simulation run. Both diversity range between 0 (homogeneity) and 1 (maximal diversity). "Diversity" is the number of distinct opinions divided by $N$, where two opinions are distinct when they differ by more than 0.005. "VarianceS" measures the standard deviation of the opinions across all population members, divided by 2 for normalization. Figure 1 clearly shows that both measures of diversity rapidly decline over time and eventually drop to zero. This result turns out to be robust when more replications and longer runs are used. Based on 100 replications, we found a mean Diversity and varianceS after 10000 iterations of virtually zero.

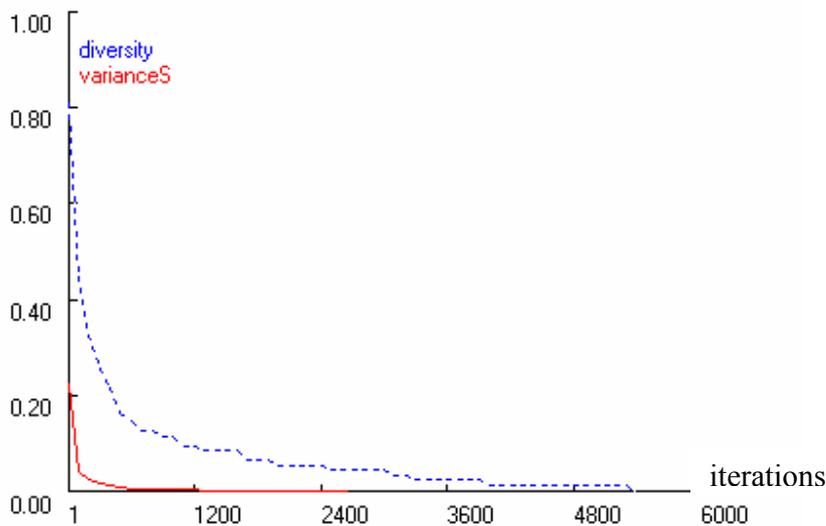

Figure 1. Change of diversity measures in one replication of experiment 1.

The result of the first experiment can be readily explained by insights from social influence models in the tradition of Abelson (1964). It has even been shown analytically in this literature that with a continuous opinion scale, a very weak condition is sufficient to guarantee full homogeneity of opinions in equilibrium: The network needs to be "compact," such that there are no subgroups that are entirely cut off from outside influences (Abelson, 1964). These models assume that the network of influence is exogenous and static over time. However, the result readily generalizes to our analysis. Except for the unlikely case that two agents adopt opposite extremes of the opinion scale from the outset, weights will always be positive and thus the network fully connected.

Experiment 2

Macy et al (2003) included heterophobia and rejection, but they have not used a continuous opinion space. This is what we do in the second experiment using the baseline condition of the first experiment. The only difference is that we now allow weights to vary between -1 and +1, introducing heterophobia and rejection. Technically, weights are now calculated according to equation 2 rather than 2a and initial weights are again set proportional to the level of initial (dis)agreement, now between -1 and +1. To measure polarization, we use the variance of pairwise agreement across all pairs of agents in the population, where agreement is ranging between -1 (total disagreement) and +1 (full agreement), measured as one minus the distance of opinions. This measure obviously adopts its lowest level of zero in a uniform opinion distribution. Its maximum level of one is obtained when the population is equally divided between the opposite ends of the opinion scale[3] at -1 and +1. With uniformely distributed opinions, the polarization measure would be approximately 0.22. Figure 3 shows the dynamics that we typically obtained in simulation runs under the conditions of experiment 2.

---

[3] To see this: In 50% of all dyads the agreement is 1, in 50% it is -1. The average level of disagreement is zero and the squared average distance, i.e. the variance, from this is 1.

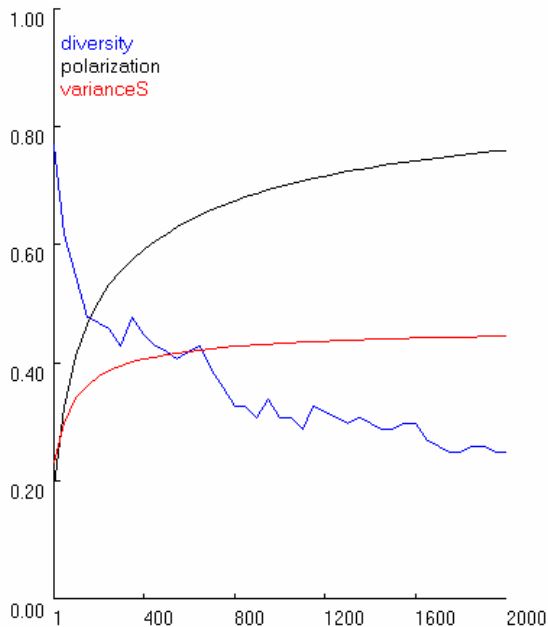

Figure 3. Change of diversity and polarization in one replication of experiment 2.

Figure 3 demonstrates that heterophobia and negative influence radically changed model dynamics as compared to experiment 1. While diversity in terms of the number of different opinions still declines over time, the variance measure now increases up to approximately 0.5. Polarization increases from the level of a uniform distribution (0.22) up to about 0.75, indicating that most group members have moved towards one of the two extreme ends of the opinion spectrum. The discrepancy between the diversity measures and polarization highlights that Axelrod's identification of stable diversity with polarization is inadequate in a continuous framework. An inspection of the corresponding opinion distribution in iteration 2000 further illustrates this point. About 90% of all population members are roughly evenly distributed between the opinions of -1 and +1, while the remaining 10% are approximately uniformely distributed along the opinion spectrum in between these two extremes. For reliability, we replicated this experiments 10 times. The average level of polarization we obtained in iteration 5000 was about 0.6, well above that of the near consensus we found in experiment 1.

The explanation for the strong tendency towards polarization in experiment 2 is that with all mechanisms operating simultaneously, there is both increasing convergence in opinions and liking (positive ties) between initially relatively similar agents and increasing difference and disliking (negative ties) between initially relatively dissimilar agents. With only one direction to move away from their enemies, agents necessarily adapt their opinions towards the extreme ends of the opinion spectrum, pulling their friends with them, or turning friendships into hostile relationships in the process. At the same time, some diversity remains due to the highly localized contact structures that we generated with a contact range of $r$=1%. With influence from only one agent to the left and right of every focal agent, it is relatively easy for the system to "discover" an equilibrium in which opposite influences from the two neighbors exactly balance so that an interior position of the focal agent remains stable. This happens when the left and the right neighbor differ from the focal agent in opposite directions by the same distance, but are both sufficiently similar to still be "friends" (positive weights). Two agents who have no contact with each other but who would be enemies (distance larger than one) if they had contact, can then still be related to each other by a chain of connected agents who each are friends of their local neighbors, but whose positions gradually shift from the one extreme position to the other. This explanation suggests that for structures with a larger contact range, it may be more difficult for the dynamics of the system to find from a random start a configuration of positions in which intermediate opinions between the two extremes can remain stable. For structures with a larger contact range, a larger number of agents needs to be coordinated in every local neighborhood. This makes it less likely that perfect polarization can be avoided from a random start.

Experiment 3

To test effects of contact range, we used the parameter settings of experiment 2 and varied the contact range from $r$=2% up to $r$=50% in steps of 2%. Outcome measures were calculated for the opinion distribution of iteration 1000, based on mean values of ten replications per condition. Figure 4 shows the results.

The figure confirms that increasing contact range $r$ entails higher levels of polarization. The polarization measure rises from a level of about 0.6 at $r$=2%, up to nearly the level of perfect polarization (1.0) from $r$=10% on. This is clearly inconsistent with results from previous models of cultural dynamics. More contact also reduces diversity. Only 2 distinct opinions (+1 and -1) survive beyond $r$=10% with a variance measure of 0.5.

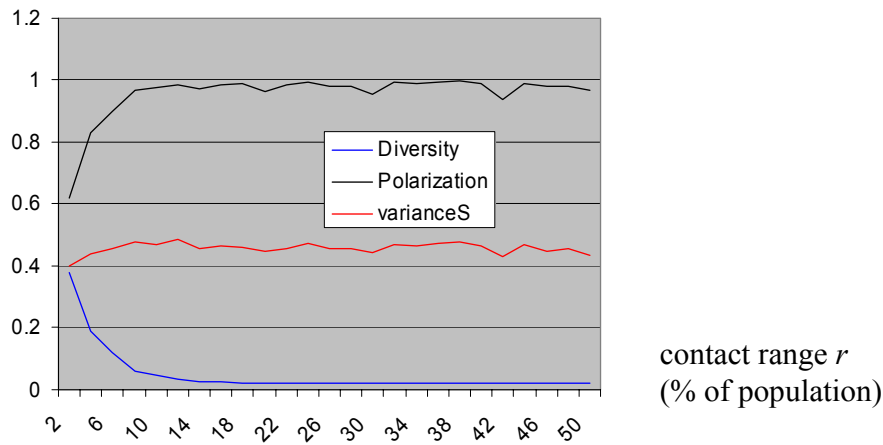

Figure 4. Results of experiment 3. Effect of contact range r on diversity measures and polarization measure in iteration 1000. Mean values of 10 replications per condition, based on parameter setting of experiment 2.

Experiment 4.

The baseline scenario of experiment 1 would have generated a high level of diversity in Axelrod's framework. But some conditions of this baseline scenario may weaken the effect of contact on polarization. To test robustness, we varied the number of issues, $k$, and the level of moderation, $m$, in experiment 4a and 4b, respectively.

In line with Macy et al (2003), we may interpret the number of issues, *k*, as an indicator of the "broad mindedness" of a society. Intuitively, broad mindedness may reduce tendencies towards polarization, because it generates more possibilities for overlap and thus for positive influence between neighbors. This corresponds to results obtained by Axelrod (1997; cf. Klemm, 2003a,b). To test this possibility, we replicated experiment 3 with *k*=2. Figure 5 shows the results of this analysis.

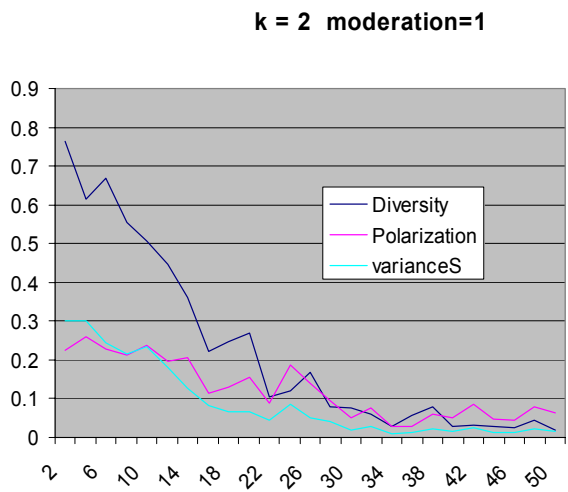

Figure 5. Results of experiment 4a. Effect of contact range r on diversity measures and polarization measure in iteration 1000 with *k*=2 issues and moderation *m*=1.

Figure 5 shows that *k*=2 radically changes the effect of contact range. While we found with *k*=1 that contact range increases polarization, we now obtain a negative effect on polarization. The explanation is that more issues reduces the likelihood of initial negative, because the probability declines that two agents are sufficiently different on all issues. As a consequence, most agents are drawn towards the average opinion in their neighborhood and thus move away from the extreme ends of the opinion distribution. A larger contact range only amplifies this convergence to the mean, because large neighborhoods are more representative of the overall population.

Consistently, we found even lower levels of polarization when the number of issues was further increased.

A higher level of moderation resurrects the effect of contact on polarization at least partially. This may seem counterintuitive at first. However, higher moderation implies that agents are influenced less by others who have similar opinions, in comparison with the influence experienced from dissimilar others. Higher moderation thus makes negative influences stronger in comparison with positive influences. We conducted a further simulation experiment (4b) in which we used the conditions of experiment 4a with one change, we increased the level of moderation from $m=1$ to $m=2$.[4]

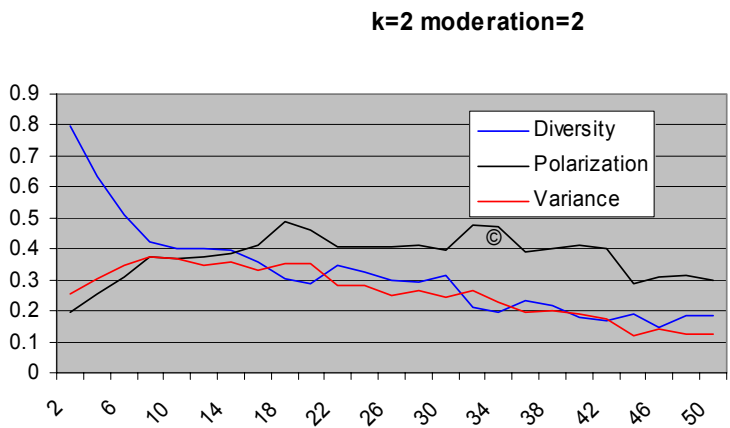

Figure 6. Results of experiment 4b. Effect of contact range r on diversity measures and polarization measure in iteration 1000 with $k=2$ issues and moderation $m=2$.

Figure 6 confirms the expectation. With higher moderation, contact range can again increase the level of cultural polarization. But the figure shows a more complex, non-linear effect than we found before for $k=1$. Higher contact range increases polarization only between $r=2\%$ and approximately $r=18\%$. The level of polarization does not change remarkably between $r=18\%$ and

---

[4] We also tested the effect of increasing moderation for the baseline condition of a single issue space ($k=1$). As suggested by the above reasoning, this only increased the already high level of polarization shown by figure 4.

*r*=42% and it declines beyond *r*=42%. This reflects the interplay of two opposite effects of contact range, increasing complexity of coordination (fostering polarization) and increasing exposure to the population mean (fostering homogeneity). Further simulations with more issues confirmed the inverted U-shaped effect, but only when moderation was likewise increased. The more broad minded our simulated society, the higher is the level of moderation at which the effect of contact range on polarization turns from a negative effect into a positive one.

4. Discussion and conclusion

Axelrod (1997) initiated prominently a line of computational models of cultural influence that suggest that more contact between geographically dissimilar regions may reduces. In this paper, we have shown that this conclusion critically hinges upon the representation of opinion scales as nominal. We showed that with the assumption of continuous opinions, stable cultural diversity can no longer be readily explained in this framework, unless negative social influence and heterophobia are added to the model. These mechanisms imply contrary to previous work, that contact may under certain conditions increase rather than decrease polarization.

This study relies on a number of simplifications that need to be carefully examined. We will discuss a number of those simplifications below and give arguments why we expect that they do not fundamentally limit generality of our conclusions. We have partially also tested these argument with explorative simulation studies and will extend this in future work. For one thing, our framework is not fully compatible with the one used in the tradition of Axelrod (1997), because we use a one dimensional geographical space rather than the two dimensional space used by Axelrod. However, analytical work by Klemm and others (2003a,b) has shown that this simplification is not critical for the results that Axelrod has obtained, so we see no apparent reason why it should be critical for ours. These analytical studies have also revealed that Axelrod's result of stable diversity may critically hinge upon another condition, the absence of random noise in opinions. We expect that noise is less critical for our main conclusions, because the polarized outcomes we are interested are highly stable against a small level of random changes in opinions. When in a polarized outcome some agents change from time to time their opinion on some issue, they are either immediately "drawn back" to the pole of their "friends", or they change sides and move to the other pole if the opinion change was large enough. A further potential limitation is that we only used a regular lattice structure and not supposedly more realistic models of social contact structures, like the celebrated small world or scale free networks. Again, we do not believe that the effects of contact range may radically different in such structures. Consider a small world network consisting of sparsely connected local clusters. Higher contact range in such a network means more ties between clusters. This has exactly the same effects on the dynamics of social influence and social selection that we found in a regular framework. It increases the exposure of all society members to the overall mean – thus hampering polarization – and it increases the coordination complexity of diverse equilibria in strongly clustered network – thus fostering polarization. As a final simplification, we have neglected that agents may not only differ on "flexible" opinions but also on "fixed" demographical characteristics (see also Macy et al, 2003). Demographical differences tend to correlate with geographical location. As a consequence, we expect that including fixed agent attributes increases rather than decreases the potential for heterophibia and rejection in the simulated societies. Hence, this complication may not limit but actually strengthen the effect of contact range on polarization.

To conclude, we believe to have demonstrated that contrary to what previous models of cultural dynamics suggest, an increase of contact between geographically distant regions may increase rather than decrease cultural polarization in society. More theoretical analyses and empirical work are needed to assess the empirical relevance of this result. In any case, our finding should caution modelers of cultural dynamics against overestimating the positive effects of contacts on cultural convergence.